\documentclass[12pt]{article}

\usepackage{amssymb,amsmath,mathbbol,mathrsfs}
\usepackage[usenames,dvipsnames]{color}
\usepackage{hyperref}
\usepackage{xcolor}
\usepackage{stmaryrd}
\usepackage{authblk}
\usepackage{framed}
\usepackage{empheq}
\usepackage{slashed}
\usepackage{hyphenat}
\usepackage{cite}

\usepackage{marginnote}    

\usepackage[left=1in,right=1in,top=1in,bottom=1in]{geometry}                  

\usepackage{sectsty}
\allsectionsfont{\sffamily\mdseries\upshape} 
\usepackage{tocloft}

\makeatletter
\renewenvironment{abstract}{%
    \if@twocolumn
      \section*{\abstractname}%
    \else 
      \begin{center}%
        {\bfseries\sffamily\abstractname\vspace{\z@}}
      \end{center}%
      \quotation
    \fi}
    {\if@twocolumn\else\endquotation\fi}
\makeatother

\numberwithin{equation}{section}
5

\hypersetup{
	colorlinks=true,         
	linkcolor=MidnightBlue,          
	citecolor=BrickRed,        
	urlcolor=MidnightBlue            
}

\newcommand\mathC{\mkern1mu\raise2.2pt\hbox{$\scriptscriptstyle|$}
        {\mkern-7mu\rm C}}              

\newcommand{\be}{\begin{equation}}
\newcommand{\ee}{\end{equation}}

\newcommand{\cint}{{\int\kern-.87em{<}}}
\newcommand{\sint}{{\int\kern-.75em{\sim}}}
\newcommand{\fint}{{\int\kern-1.00em{\int}}}

\let\oldmarginpar\marginpar
\renewcommand\marginpar[1]{\oldmarginpar{\color{red}\raggedright\footnotesize #1}}

\title{John Bell on ‘Subject and Object’: an Exchange }
\author{Hans Halvorson and Jeremy Butterfield  
}

\begin{document}

\maketitle

\begin{abstract} This three-part paper comprises: (i) a critique by
  Halvorson of Bell’s (1973) paper ‘Subject and Object’; (ii) a
  comment by Butterfield; (iii) a reply by Halvorson. An Appendix
  gives the passage from Bell that is the focus of Halvorson's
  critique. \end{abstract}

\bigskip 
\begin{center}  
\large{Part I: John Bell on Subject and Object; by H. Halvorson}
\end{center}

It's quite amazing that in the span of four short pages, John Bell can
make the pioneers of quantum mechanics seem collectively like just so
many addle-brains.  I'm speaking here of Bell's article ``Subject and
Object''.\footnote{The paper was presented at a symposium in honour of
  Dirac in September 1972, published in {\em The Physicist’s
    Conception of Nature} (ed. J. Mehra) in 1973, and reprinted in
  Bell’s collection (1987; second edition 2004, with the same
  pagination). Compare the Appendix.}  I cannot deny the rhetorical
effectiveness of this article.  In fact, I consider it a model for how
one can --- with the effective application of insinuation and
rhetorical question --- render a view seemingly unworthy of serious
consideration.  Nonetheless, I cannot hold Bell's paper up as a
paradigm of philosophical inquiry, because he gives so little effort
to understanding what others were saying.  We can do better, and we
must do better, if we're ever going to make progress with the
foundations of quantum physics.

Bell begins his article by claiming that:
\begin{enumerate}
\item Quantum mechanics is fundamentally about the results of
  ``measurements''.
    \item The subject-object
      distinction is needed for quantum mechanics, but
    \item ``Exactly \emph{where} or \emph{when} to make it [i.e. the
      subject-object distinction] is not prescribed.'' (p 40)
    \end{enumerate}
    \noindent Bell then says that (3) is a serious defect that makes
    quantum mechanics ``vague'' and ``intrinsically ambiguous'' and
    ``only approximately self-consistent.'' (We have included the
    complete text of the relevant passage in an appendix.)

Let me begin by saying that I simply deny (1), i.e.\ that quantum
mechanics is fundamentally about the results of measurements.  I'm
afraid that Bell has himself made a logical leap from ``the quantum
mechanical formalism needs a user'' to ``quantum mechanics is
fundamentally about the results of measurements.''  There is a wide
range of possibilities between these two extremes --- e.g.\ that the
quantum-mechanical formalism provides a means for translating facts
about subatomic reality into a language that human beings can
understand.

I will grant that Bell is correct about (2), that the subject-object
distinction is needed for quantum mechanics, but unfortunately, Bell
has misunderstood the sense in which it is needed.  He seems to think
that quantum mechanics must describe the world as bifurcated into two
parts --- subject and object.  If that were correct, then I would
completely understand Bell's unease with the distinction.  If the
theory describes a world with two parts, then the theory should offer
some guidance about what belongs to each part.

But if you think about the meaning the word ``subject'', it quickly
becomes obvious that it's not supposed to play the role of a predicate
in the theory (unlike, say, ``electron'').  Rather, the idea is that a
subject uses the theory to describe objects --- and in the case at
hand, these objects fall under the laws of quantum mechanics.  The
theory sees no subjects, it sees only objects, and so it has no need
for specifying \emph{where} and \emph{when} the subject-object split
occurs.  Such a split is a necessary prerequisite to physical
theorizing, when a subject decides to use a theory to try to say
something true about the world.


Now what about the complaint that quantum mechanics does not specify
who the subject is, or when and where and how she decides to use the
theory?  But wait a minute.  Is there any theory that does that?  What
an amazing theory it would be!  Indeed, such a theory would fulfill
Hegel's aspiration of finally unifying the subject and object.  In
other words, such a theory would ``theorize itself.''  Is Bell
suggesting that quantum mechanics is defective because it doesn't yet
achieve the Hegelian \emph{Aufhebung} of the subject-object
distinction?

So, in short, Bell is correct that quantum mechanics, as it stands,
needs a subject.  But that is true of every theory that has ever
appeared in physics --- i.e.\ these theories need subjects to decide
when and where and how to describe things.

Bell's subsequent rhetoric in the article is effective only against
the backdrop of his false assumption that the subject must appear
\emph{in} the quantum-mechanical description.  For example, Bell
raises a question for which quantum mechanics doesn't appear to have
an answer.
\begin{quote} Now must this subject include a person?  Or was there
  already some such subject-object distinction before the appearance
  of life in the universe? (p 40) \end{quote} But quantum mechanics is
simply not interested in the question of what counts as a subject.  If
you ask me what counts as a subject, then my answer is that anyone who
can use a theory to describe things is a subject --- no other
qualifications are necessary!  If your dog can theorize, then he is a
subject, and if an artificial intelligence could theorize, then it
would also be a subject.  And to Bell's second question, I suspect
that before the appearance of ``life'' in the universe, there were no
things that could describe other things, and hence no subjects.  But
that doesn't mean that we subjects, living today, cannot describe the
universe as it was before the existence of any subjects.  In fact, the
entire point of the subject-object distinction is that when a subject
$S$ is treating some $X$ as an object, then it is indifferent to $S$
whether $X$ is also a subject --- because as far as $S$ is concerned,
$X$ is merely an object.

If you now ask me, but is $X$ \emph{really} a subject or an object?
Here I say that the question is misguided.  Those two categories are
not mutually exclusive.  Without a doubt, each subject in our world
can be an object of some subject's description.  So perhaps what you
want is a more comprehensive theory that answers the question of who
or what can be a subject.  But then who would be the subject who uses
that theory, and must she wait for the theory to tell her that she is
a subject before she can make use of it?  I feel that we have now swum
into deep metaphysical waters.  For the business of physics, is it not
enough that the subjects know who they are?

Due to misunderstanding the role of the subject in quantum mechanics,
Bell also falsely accuses quantum mechanics of being
``\emph{intrinsically} ambiguous and approximate'' (p 41, emphasis in
original).  If quantum mechanics does not describe a world split into
subject and object, then where is the ambiguity supposed to appear?
If Bell says that the ambiguity arises in what quantum mechanics is
intended to describe --- i.e.\ what counts as the object --- then I
would ask how that is different from any other physical theory.  Take
one of Bell's favorite theories: Bohmian mechanics.  What is Bohmian
mechanics supposed to describe?  You might say: it describes particles
following deterministic trajectories.  But then I would ask: which
particles, and which trajectories?  You see, even in Bohmian
mechanics, it's left to the discrimination of the theoretical
physicist to decide how many particles, which Hamiltonian, when the
interaction turns on and off, etc.\footnote{Consider, for example, the
  Bohmian description of a momentum measurement: According to Norsen,
  ``one could `turn off' the potential energy $V(x)$ which confines
  the electron to the vicinity of the origin \dots '' (Norsen 2017, p
  196).  To echo Bell's question, exactly \emph{where} and \emph{when}
  is the potential energy turned off?}  So, if standard quantum
mechanics is ``\emph{intrinsically} ambiguous and approximate'' how is
that not also the case for Bohmian mechanics?

In ``Subject and Object'', Bell slices and dices his opponent --- a
straw person of Bell's own making.  The real problem, I think, is that
Bell wants a theory that has no need for a subject.

\bigskip \begin{center}
\large{Part II: Comment on Halvorson;  by J. Butterfield}
\end{center}  

\section{Introduction}\label{intro}

In this Comment,  I will maintain that Halvorson's criticisms are unfair to Bell. As an introduction, let me begin with the charges in Halvorson's opening paragraph: that Bell made `the pioneers of quantum mechanics seem ... like ... addle-brains', and that he gave `so little effort to understanding what others were saying'. I maintain that Bell is innocent of the charges.  My reasons, in short, are that:\\
\indent \indent (i): Bell's discussion in this paper of `subject' and `object'---meaning, as he explains, `measurer' and `measured system'---is a brief statement of the measurement problem; as is announced by Bell's opening sentence: `The subject-object distinction is indeed at the very root of the unease that many people still feel in connection with quantum mechanics' (1987, p. 40).  \\
\indent \indent (ii): Bell does not shirk his duty to try to
understand what the pioneers of quantum mechanics said about the
measurement problem. On the contrary, he starts his third paragraph by
saying that they were aware of these questions ...
\begin{quote} ... but quite rightly did not wait for agreed answers
  before developing the theory. They were entirely justified by
  results. The vagueness of the postulates in no way interferes with
  the miraculous accuracy of the calculations. Whenever necessary a
  little more of the world can be incorporated into the object. In
  extremis the subject-object division can be put somewhere at the
  `macroscopic' level, where the practical adequacy of classical
  notions makes the precise location quantitatively
  unimportant. \end{quote}

\indent \indent (iii): Besides, one must of course read this
paper---like any paper---in context. And Bell's other papers contain
longer statements of: both (a) the measurement problem, including in
the terms used here, viz. the `ambiguity' about where the `cut'
between object and subject, or `measured system' and `measurer',
should be made (which he elsewhere calls `the shifty split'); and (b)
the pioneers' exertions over, and insights about, the measurement
problem.\footnote{\label{jsbquotes}{Agreed, the papers with the
    best-known of these longer statements of (a) and (b) were written
    after `Subject and Object’: for example, `On the impossible
    pilot-wave’, `Speakable and unspeakable\dots ', `Six possible
    worlds\dots ', and `Against measurement' (Chapters 17, 18, 20 and 23 of (1987/2004)). But I thank Chris Timpson for pointing out to me that one also finds earlier statements, in `The moral aspect of quantum mechanics' (from 1966: Chapter 3), in `On the hypothesis that the Schr\"{o}dinger equation is exact' (from 1971; the revised 1981 version,  called `Quantum mechanics for cosmologists',  being Chapter 15)  and, more briefly, in Section 1 of `Introduction to the hidden variable question' (1971, Chapter 4) . Note also: (i) Bell's 1989 Trieste Lecture, which talks in detail about `the shifty split' and about Dirac (cf. Bassi and Ghirardi (2007)), and (ii) the quotes in Ghirardi's touching memoir (2014).}}   \\

But what about the substance of  Halvorson's criticism of Bell? Halvorson is very clear. His main criticism is that Bell has made a false assumption: that the subject, i.e. the measurer or measuring system,  must appear {\em in} the quantum-mechanical description. I maintain that this criticism is wrong. Bell does not assume this. What he {\em does} do---here and in other papers (cf. footnote \ref{jsbquotes})---is to contrast:\\
\indent \indent (a) what he sees as the happy situation in classical physics: that there seems no obstacle, in principle, to a classical physical description of measurement processes that successfully describes getting definite measurement results;   \\
\indent \indent  (b) what he sees as the unhappy situation in quantum physics: that there seems to be an obstacle, in principle, to a quantum physical description of measurement processes that successfully describes getting definite measurement results---this is the measurement problem.  \\

\noindent I do not mean to put Bell on a pedestal, or treat him as the fount of all wisdom about interpreting quantum mechanics. To be sure, his discussions are brilliant and his work on non-locality was, obviously, epoch-making. (And speaking for myself: his realist philosophical outlook is music to my ears.) But a good case can be made for some views that he gave short shrift to. One main example is Bohr’s doctrine of the necessity of classical concepts: whose formulation and defence has been deepened in the intervening years, notably by Halvorson himself (e.g. Halvorson and Clifton 2002) and by Landsman (2006, 2017 especially Introduction).\\

\noindent So in the rest of this Comment, I will: reply to Halvorson's criticism, by expanding on the contrast between (a) and (b) (Section \ref{us}); and conclude more positively (Section \ref{concl}). 

\section{Reply to Halvorson's criticism}\label{us}

Let us recall Halvorson's opening trio of claims that he attributes to
Bell. Halvorson writes:---
\begin{quote}
  Bell begins his article by claiming that: \\
  \indent (1):	Quantum mechanics is fundamentally about the results of `measurements'.\\
  \indent (2):	The subject-object distinction is needed for quantum mechanics, but \\
  \indent (3):	``Exactly where or when to make it [i.e. the subject-object distinction] is not prescribed.'' (p 40)\\
  Bell then says that (3) is a serious defect that makes quantum
  mechanics ``vague'' and ``intrinsically ambiguous'' and ``only
  approximately self-consistent.''

Let me begin by saying that I simply deny (1) \dots
\end{quote}

I will focus on Halvorson's discussion of his (2), and thereby
(3). For I believe the apparent Bell-Halvorson disagreement over (1)
need not detain us. For I think it is clear that for Bell, (1) has two
roles: but the first prompts no dispute and the second is covered by
the dispute over (2) and (3).

The first role of (1) is that Bell evidently intends it as a report of
the orthodox ways of thinking about quantum theory, not as his own
view. In this role, (1) just helps set up Bell's discussion. In the
second role, (1) serves to introduce measurement results as an
undeniable focus of the enterprise of physics: physics is undeniably
in the business of accounting for measurement results. Here, I say
`account for' to cover indifferently: (i) prediction (and
retrodiction), definite or probabilistic, and-or (ii) explanation,
and-or (iii) other relations of `meshing' between the claims of a
physical theory and empirical phenomena, such as confirmation.

Neither I nor Bell need to choose between these. For of course, in
this second role, (1) is introducing the measurement problem. That is:
it stresses the pervasive and detailed success of classical physics in
attributing to all objects that it is applied to, definite values for
all the quantities appropriate to them. This is often summed up in the
slogan that `measurements have definite results'; or that according to
classical physics, they do. But of course, all parties agree that the
point at issue goes far beyond measurements, and encompasses all
objects, measured and unmeasured, to which classical physics
successfully applies. Accordingly, in view of classical physics'
supreme success in describing macroscopic objects as having definite
values for all their quantities, the point is often summed up as: `the
definiteness of the macro-realm'.

And {\em this point} yields the quantum measurement problem. For there
is an argument---the familiar one: Schr\"{o}dinger's argument about a
cat!---that this point is incompatible with quantum physics. More
precisely: it is incompatible with the quantum dynamics of a strictly
isolated system being unitary. And it is no escape from this quandary
to point out that the cat (i.e. the pointer of an apparatus set to
measure a quantity on a micro-system that is in a superposition for
that quantity) is not strictly isolated, since it is interacting with,
for example, air molecules, and indeed the CMB. For the official
quantum state of the cat (or pointer), after the poisoning/measurement
process, that is obtained by tracing out its environment---although it
is mathematically a mixture---cannot be given the ignorance
interpretation. In d'Espagnat's terminology: it is an improper (not
proper, i.e.  ignorance-interpretable) mixture.\footnote{D'Espagnat
  suggested these terms in (1976: Chapter 6.2). Nowadays, the point is
  often made in the literature on decoherence (e.g. Zeh, Joos et
  al. (2003, p. 36, 43); Janssen (2008, Sections 1.2.2, 3.3.2)). But
  it is humbling to recall that the point was already clear, and
  beautifully expressed, in Schr\"{o}dinger's amazing 1935 papers:
  cf. especially the ``cat paradox” paper's analogy with a school
  examination (1935, Section 13, p. 335f.).}

Given all this; what about `subject' and `object', i.e. Halvorson's
(2) and (3)?  As I announced in Section \ref{intro}, and we read in
Halvorson's text: his main claim against Bell is that Bell falsely
assumes that the subject must appear {\em in} the quantum-mechanical
description. But I submit that Bell does not assume this. Rather, he
emphasises a contrast between quantum and classical physics. As I put
it at the end of Section \ref{intro}: in quantum physics, there seems
to be an obstacle, in principle, to a quantum physical description of
measurement processes that successfully describes getting definite
measurement results.

In other words: there is an argument (Schr\"{o}dinger's argument about
a cat) that quantum physics cannot recover, or secure, the
definiteness of the macro-realm. For a suitable `diabolical
device'---a `ridiculous case': both are Schr\"{o}dinger's phrases
(1935, p. 328)---could propagate the {\em indefiniteness} of the
micro-realm into the macro-realm.  On the other hand, within classical
physics, there seems to be no such obstacle, no such argument:
measured systems can be coupled to apparatuses, in such a way that the
definite values of their quantities can be registered by those
apparatuses' pointers.

Besides, this contrast can be `pushed inside the head', if we so
wish---and as the jargon of `subject' and `object' suggests it might
be. So far, despite talk of `measurement' with its connotations of
human activity and cognition, it is the {\em inanimate} macro-realm,
such as the definite positions of pointers, that I have
emphasised. But (notoriously!) some authors suggest we should push
`von Neumann's chain'---the successive coupling of systems,
correlating appropriate eigenstates, so as to get a many-component
entangled state (cf. von Neumann 1932, Chapter VI.1
p. 418-420)---inside the head, and thereby consider the quantum
mechanical description of the neural correlates of experience,
e.g. seeing the black pointer inclined leftward against a white
background vs. seeing the black pointer inclined rightward against a
white background. If we concur with these authors, and countenance
such a many-component entangled state as the complete physical
description, we seem to face a looming threat of `indefinite, or
superposed, experiences'. And we must choose between two broad options
for avoiding the threat, i.e. for securing definite appearances of
e.g. a pointer. That is: for securing an {\em apparently} definite
macro-realm. Either we adopt an Everettian viewpoint (in a broadly
`many-minds', rather than `many-(inanimate)-worlds', version); or we
say that `consciousness collapses the wave-function'.\footnote{Famous
  examples of these two broad options include Zeh (1970) and Wigner
  (1962), respectively. Bell's `Six possible worlds of quantum
  mechanics' (1987/2004, Chapter 20) is a breezy introduction to both
  options, among others.}

But in this paper, I of course do not need to choose between these options. For I come, not to solve the measurement problem, but only to praise it: or at least, to prevent it being buried ... Here, my point is---as it was in my discussion of the  inanimate macro-realm---that in classical physics, there seems to be no such obstacle, no such argument, against maintaining both:\\
\indent  \indent (i) all experiences being definite, and \\
\indent  \indent (ii) there being a complete physical description of the neural correlate of any experience. \\
Just think of modern psychophysics, with its reliance on
neurophysiology formulated wholly in classical-physical terms,
e.g. with stick-and-ball models of the underlying biochemical
molecules. Think in particular of Hubel and Wiesel's 1950s
investigations of vision. They found that in the visual cortex of a
cat ({\em sic}), a single specifically-located neuron is dedicated to
firing in response to an edge being aligned at a certain angle from
the vertical (say, 20 degrees, as vs. 10 or 30) in a certain region of
the visual field (say, the top-left region). Indeed, one can imagine
the edge in question being a black pointer inclined leftward against a
white background. Besides, the firing of the neuron is understood in
classical neurophysiological terms as an electrical impulse,
underpinned by sodium and potassium transport. So it fires---or it
doesn't. The cat detects the edge (the pointer) inclined leftward at
20 degrees from the vertical---or it doesn't.

This completes my Comment on Halvorson. But there are three ancillary
points that are worth making \dots

\subsection{Philosophy, history and the pilot-wave}\label{threeancillary}
The first two points are about the classical case. The first is about avoiding some philosophical commitments; the second is a historical point  about the success of science supporting philosophical materialism. The third point is about the quantum case, and `extra variables'.\\

(1): Note that our `no worries' attitude, for classical physics,  about describing the `subject', even experiences themselves, does not require: (a) the coherence of describing at once all the subjects in the cosmos; nor (b) philosophical materialism. 

As to (a), I have gestured at how, in a world described by classical
physics, `ordinary' i.e. conceptually unproblematic empirical enquiry
would be able in principle to discover the detailed physical
description of any object or event, including measurement results; and
even if one interprets `measurement' in terms of experience, nothing
in principle prevents classical physics from describing the neural
correlates of experience. But this does not commit us to saying that
in such a world, classical physics could describe it ``all in one
go''. It is of course a matter of the order of quantifiers: `for each
object and event, there could be a classical physical description'
does not imply `there could be a single classical physical
description, for all objects and events'.\footnote{As it happens, I
  have no have qualms about the sort of classical cosmic inventory
  envisaged by the last sentence. I agree, of course, that it might
  well be infinite, and so ungraspable by human minds. But I do {\em
    not} take the propositions---the descriptions, the items in the
  inventory---to be a part of the cosmos described; and so there is no
  problem of the inventory itself having to be listed, or of
  self-reference or regress.  But if you have such qualms: rest
  assured that nothing I, or Bell, have said commits one to such an
  inventory.}

As to (b): nothing I  said about neural correlates of  experience (in a world described by either classical or by quantum physics)  requires philosophical materialism. I take materialism to be a thesis of supervenience or determination.  It says, roughly speaking, that all the facts about the whole cosmos supervene on,  or are determined by, the facts as described by the natural sciences. (I here take `natural sciences' to encompass physics, chemistry and biology, on a par, i.e. with no special status accorded to physics: I will return to this in (2).)   But I will not need to pursue a precise formulation of materialism. For me, the main point is that materialism is meant to {\em exclude} all non-natural-scientific properties and relations, even ones that are strictly nomologically correlated with some natural-scientific property or relation. Such  properties and relations---non-natural-scientific but nomologically correlated  with the natural-scientific---are invoked by some traditional {\em anti}-materialist views, like epiphenomenalism and property-dualism. So the point here is that what I said about the neural correlates of  experience does not exclude such properties, or such views. This leads in to (2).  \\

(2): Notwithstanding my liberal tolerance, in (1), of  epiphenomenalism:--- Consider the vast success since about 1850 of the natural sciences, i.e. physics, chemistry and biology, in describing and explaining phenomena, including mental phenomena.  Think of the rise of physiology and psychophysics (in the mid-nineteenth century: figures like Bernard and Helmholtz), the decline of vitalism in biology, the rise of biochemistry and molecular biology. And think of how physics has provided an ever more detailed underpinning of chemical and biological phenomena (and so also, it seems: of mental phenomena). 

These developments have undoubtedly prompted philosophers to formulate
philosophical materialism; and also prompted many of them to defend
the doctrine, thus formulated. (Of course, that is as it should be:
positions debated in academic philosophy should reflect---make
precise, and improve!---currents in the wider intellectual culture.)
And hence, my unblushing statement a few paragraphs above of my main
point. Namely: in classical physics, there seems to be no obstacle, no
argument, against (i) all experiences being definite, and (ii) there
being a complete physical description of the neural correlate of any
experience.

I said it unblushingly, precisely because of the rampant success of
classical neurophysiology. In 1850, or even in 1900, it could not have
come so trippingly off the tongue.\footnote{I surmise that von Neumann
  was articulating the same confidence about the conceptually
  unproblematic status of classical psychophysics when in his (1932:
  Chapter VI.1 p. 419) discussion of measurement, he talked about the
  `psycho-physical parallelism' being undercut by a {\em quantum}
  measurement: a confidence that, I say, was by 1932 well
  warranted. The point is also familiar in the history of analytical
  philosophy of mind. Recall McLaughlin's much-cited account (1992) of
  how `British emergentism', represented by e.g. C.D. Broad in the
  mid-1920s, declined not least because the rapid successes of quantum
  chemistry (e.g. London and Heitler's work on covalent bonding)
  undercut Broad's conjectured configurational forces.

  Recall also Shimony's vivid phrase, {\em `closing the circle'}, for
  the endeavour of recovering the `manifest image' of the
  world---including the definite macro-realm, or at least definite
  appearances---from the `scientific image' of it. In these terms, my
  point is that closing the circle seems easier in a world described
  by classical physics, rather than by quantum
  physics. \label{shimony}}

(3): In my Comment on Halvorson---my Bellian attempt to prevent the
measurement problem being buried---I assumed throughout that an
appropriate many-component entangled state was the complete physical
description to be considered. For example, when I `pushed the
subject-object distinction inside the head', it was this assumption
that led to the looming threat of indefinite, or superposed,
experiences. Of course, many advocates of proposed solutions to the
measurement problem will deny this assumption, and announce this
denial as their first step on the road towards their preferred
solution. I of course say: `More power to you, and good luck, in
developing your preferred solution. I have no brief to defend the
assumption---I only made it, so as to better locate Bell's dialectical
position in 1973, and to defend him as innocent of the charges laid
against him'.

But Halvorson's mention of the pilot-wave (in the penultimate
paragraph of Part I) prompts a final comment. Halvorson stresses that
for the pilot-wave theory, as for any physical theory including
orthodox quantum theory, an application of it focusses on a part of
the world, leaving other features, such as the specification of the
potential to which the quantum system is subjected, as an `external'
issue, `put in by hand', or `up to the theorist or experimentalist'.

With which I agree: indeed so. But I---and the pilot-wave
theorist---then add that this similarity between the pilot-wave theory
and orthodox quantum theory (and indeed any physical theory) is
neither here nor there. For a solution to the measurement
problem---whether the pilot-wave solution or another---in no way needs
to deny this innocuous role of an `external subject'. What matters is
to have---which the pilot-wave theorist claims to have---a solution to
the measurement problem: facts that secure a definite
macro-realm---for example (in the simplest and most familiar version
of the pilot-wave theory), the definite positions of
point-particles.\footnote{\label{2suppl}{Two supplementary
    comments. (1) Having broached the topic of the neural correlates
    of definite experiences, I should note a misgiving about this
    solution. Even if one accepts that the definiteness of the
    inanimate macro-realm is a matter of point-particles' positions
    being {\em here} rather than {\em there}, the pilot-wave theory,
    in order to secure our having definite experiences, presumably
    requires that an experience being definite—one way rather than
    another—involves point-particles being in one wave-packet rather
    than another. But that seems hard to line up with, for example, an
    edge-detector cell in a cat's visual cortex either firing or
    not. For discussion, cf. e.g. Brown and Wallace (2005: Section 7,
    p. 533--537).

    (2): I thank Ronnie Hermens for pointing out that Halvorson’s
    stressing that any theory `needs users' (i.e. leaves features
    outside the described system as `external' and `up to the
    physicist') is echoed in the non-locality literature, especially
    in the wake of the Jarrett-Shimony distinction between parameter
    independence and outcome independence. In particular, Seevinck and
    Uffink make explicit the different theoretical roles of
    apparatus-settings and outcomes, when they write `to specify how
    probable it is that Alice will choose one setting [] rather than
    [another \dots ] would be a remarkable feat for any physical
    theory. Even quantum mechanics leaves the question what
    measurement is going to be performed on a system as one that is
    decided {\em outside} the theory, and does not specify how much
    more probable one measurement is than another' (2011, Section
    III.B). I would add that besides, one can `shift the split' i.e.`
    move the cut'. That is: one can instead model an
    apparatus-setting, and the act of choosing a setting, as a
    deterministic function of the state of the world, and then recover
    Bell’s theorem, and cousins like the Free Will theorem, by
    assuming these functions have suitable kinds of independence
    (Cator and Landsman 2014; especially pp. 784-786; Landsman 2017a,
    especially pp. 101-102). And for a recent judicious defence of the
    idea that setting {\em dependence} is tenable, I recommend Hermens
    (2019).}}
   
\section{Conclusion}\label{concl}
I rest my case: I urge that Bell is innocent of the charges laid
against him. But let me end on a more positive and forward-looking
note. There is a major part of Bell’s paper `Subject and Object' that
neither Halvorson nor I have touched on. But I should do so, since it
answers Halvorson's clarion-call in his first paragraph (quoted in
Section \ref{intro}) that `\ldots we must do better, if we're ever
going to make progress with the foundations of quantum physics'.

After urging the measurement problem in the way that Halvorson has criticised and I have defended, Bell  goes on to:\\
\indent \indent (i) state his distinction between `observable' and {\em beable} (so far as I know, this is the first paper to advocate the jargon of `beables'); \\
\indent \indent (ii) sketch how one might formulate a Lorentz-invariant quantum theory, in which a select subset of observables (i.e. conventional physical quantities) are promoted to be beables.\\
Thus he writes
\begin{quote}
  Many people must have thought along the following lines. Could one
  not just promote some of the `observables' of the present quantum
  theory to the status of beables? The beables would then be
  represented by linear operators in the state space. [footnote
  suppressed] The values which they are allowed to {\em be} would be
  the eigenvalues of those operators. For the general state the
  probability of a beable {\em being} a particular value would be
  calculated just as was formerly calculated the probability of {\em
    observing} that value. The proposition about the jump of state
  consequent on measurement could be replaced by: when a particular
  value is attributed to a beable, the state of the system reduces to
  a corresponding eigenstate. It is the main object of this note to
  set down some remarks on this programme. Perhaps it is only because
  they are quite trivial that I have not seen them set down already.
\end{quote}
... and so on! 

Tempting though I find it to quote the page-long sketch that follows
(and that concludes Bell's paper), I will forebear. Suffice it to say
that the sketch exemplifies precisely the line of thought that led to
various later efforts to formulate a Lorentz-invariant,
``no-collapse'' but ``one-world'', quantum theory. These efforts are
many and varied. They include of course work by Bell himself; but also
work since Bell's death, for example on the modal interpretation---and
by Halvorson himself, such as Halvorson and Clifton (1999). And the
tradition continues: for example, in Kent's recent proposals (2014,
2015, 2017).

This is not the place to report details of these efforts.\footnote{For
  example, Butterfield and Marsh (2018, 2019) discuss Kent's
  proposals.} But I mention them (along with, of course, some of the
other work cited above, e.g. in footnote \ref{2suppl}) in order to
convey a positive message to Halvorson, and to the reader: we {\em
  can} make progress with the foundations of quantum physics.
      
\newpage 
\begin{center}  
\large{Part III: Reply: by H. Halvorson}
\end{center}

\setcounter{section}{0}

\section{Epistemological Lesson $\geq$ Measurement Problem}

There is much to thank John Bell for. He is responsible, to a great
degree, for the rebirth of the foundations of physics that occurred in
the 1960s, after the more pragmatic period around World War
II. Nonetheless, historians and philosophers of science have an
important responsibility to contextualize and assess Bell's
contributions, and to point out cases, if there be any, where he got
things wrong. It was to that end that I attempted to initiate a
discussion with a brief critique of Bell's ``Subject and Object.''
I'm grateful that Jeremy Butterfield joined the discussion, and I hope
that it will further clarify Bell's role in physics and its
philosophy.

Let me come straight to the point. I am in complete agreement with
Butterfield about the value of clear, rigorous, and honest thinking
about the foundations of physics. In particular, we should not turn a
blind eye to genuine foundational problems. If Butterfield and I have
any disagreement, then, it is primarily about where to focus our
efforts. While Butterfield wants to keep the measurement problem
alive, I want to resurrect the ``epistemological lesson'' of quantum
mechanics that Niels Bohr spent forty years trying to articulate (see
e.g.\ Bohr 1928), and John Bell spent thirty years trying to bury.

Bell's smear campaign against Bohr has been far more successful than
he could have dreamed: these days it is all too easy to find articles
and books --- even bestsellers! --- where Bohr is described as
dogmatic, unphilosophical, obfuscating, uninterested in questions of
reality, etc. And it is increasingly difficult to find literature
where Bohr is treated as an important thinker (as Einstein treated
him). I will make three claims about this sad state of affairs, where
propaganda has prevailed over accuracy: First, these kind of slurs
about a person's intellectual character have no place in serious
philosophical discussion. Second, the objective historical evidence
gives a rather different picture of Bohr than the one Bell painted, a
picture of a person deeply engaged with questions about reality, but
without being dogmatic that he had the answers.\footnote{Just a couple
  of examples: Einstein in a 1954 letter: ``[Bohr] utters his opinions
  like one perpetually groping and never like one who believes he is
  in possession of definite truth.''  Schr\"odinger in a 1926 letter:
  ``There will hardly again be a man who has achieved such enormous
  external and internal success, who in his sphere of work is honored
  almost like a demigod by the whole world, and yet who remains --- I
  would not say modest and free of conceit --- but rather shy and
  diffident like a theology student. \ldots this attitude works
  strongly sympathetically in comparison with the excessive
  self-confidence that one often finds in the medium-sized stars of
  our profession. \ldots [Bohr] is so very considerate and is
  constantly held back by the fear that others could take an
  unreserved assertion of his (i.e.\ Bohr's) point of view as an
  insufficient recognition of others' (in this case, my)
  contributions.'' (translation by H.H.)} Third, and most relevantly,
this trash-talking of Bohr begins with the polemical works of John
Bell --- a man who never once spoke with Bohr, and who had little
knowledge of Bohr's intellectual context.\footnote{I deduce that Bell
  never spoke with Bohr from his 1988 \emph{Omni} interview, where he
  mentions that he was once in an elevator with Bohr, but did not work
  up the nerve to speak to him. For details about Bell's background,
  see (Whitaker 1998, 2016).} The picture of Bohr as dogmatist,
obscurantist, intellectual bully, etc.\ is a creative fiction of John
Bell.

\section{Bell in context}

Let me now confess my own scholarly sins. First, as Butterfield
correctly points out, I failed to contextualize ``Subject and Object''
in Bell's larger corpus, and in particular, I failed to point out how
it fits with the theme of ``there is a quantum measurement problem.''
The lack of contextualization was a result of the brevity of my
piece. However, I maintain that a similar critique can be maintained
and even strengthened against the backdrop of Bell's entire corpus. In
particular, Bell gets top marks for persuasive rhetoric, but low marks
for historical accuracy (perhaps on purpose). What's more, while Bell
was good at drawing out the absurd implications of the views of some
practicing physicists, he did not inquire into the presuppositions of
his own views. In fact, I believe that if Bell were pushed to clarify
his own philosophical commitments, then he would eventually have to
admit that he demands a physical theory to describe things ``as they
are in themselves'' (to use a phrase from Bernard Williams). This kind
of commitment sounds really good, until one starts asking hard
questions about the presupposed metaphysical and semantic picture. For
example, does a theory describe, or is it a human subject who
describes?  And if a human subject is doing the describing, then how
could he or she describe without employing arbitrary elements, such as
specific linguistic conventions?\footnote{In footnote \ref{shimony},
  Butterfield mentions Shimony's project of ``closing the circle'',
  i.e.\ deriving the manifest image from the scientific image. If the
  ``scientific image'' is things as they are in themselves, and the
  ``manifest image'' is things as they appear to us, then closing the
  circle would amount to achieving Bernard Williams' ``absolute
  conception of reality''. This noble aspiration was critiqued by,
  among others, Putnam (1992).} Such questions make it seem a little
less obvious that a physical theory can and should describe reality as
it is in itself.

Second, I admit that I should have explicitly noted that ``Subject and
Object'' was addressed to a particular audience at a particular time
--- and that it was a time when physics was in a bad state,
philosophically speaking. Bell mentions elsewhere that his university
courses in quantum physics were philosophically frustrating (see
Whitaker 1998); and surely he was acquainted with many physicists who
did not care about foundational rigor. So, I grant that Bell did have
a legitimate complaint.

Unfortunately, it did not occur to Bell that originally fruitful ideas
might have become corrupted by less philosophically reflective
physicists. Indeed, while Bell correctly identified a problem in how
people were approaching physics, he incorrectly diagnosed the nature
and the sources of the problem. The problem with physics was not
Bohr's influence, and especially not Bohr's attempts to reflect on
physics at a deep philosophical level. Rather, the problem was that
physics was increasingly being pressed into the service of
technological and military dominance, with the result that physics
education came to focus more on technique and calculation than on
conceptual understanding. The injunction to ``shut up and calculate''
did not emerge from pre-war Copenhagen, but from the rising
technocratic superpower, the United States.\footnote{To properly
  support these general claims, it would be useful to look at funding
  trends, or changes in physics curricula and textbooks.}

It might seem like I am trying to direct attention away from the real
problem, which is simply the measurement problem. In one sense, yes, I
am intentionally trying to take in the bigger picture, and to loosen
obsessive focus on analytically tight presentations of the measurement
problem. On the face of it, the measurement problem says that a few
propositions, say $\phi _1,\phi _2,\phi _3$, are inconsistent, and so
one must reject at least one of these propositions. But of course,
$\phi _1,\phi _2,\phi _3$ are not formally inconsistent; they are
inconsistent in the light of innumerable tacit background assumptions,
e.g.\ about how we use mathematical objects to represent physical
states of affairs. What this means is that one may legitimately refuse
to play the game of saying whether one rejects $\phi _1,\phi _2$ or
$\phi _3$. For example, the typical way of trying to reduce the
Everett interpretation to absurdity is by saying that it rejects the
claim that measurements have outcomes. But Everettians rightly tell a
more sophisticated story about their solution to the measurement
problem (see Brown and Wallace 2005).

Butterfield suggests that criticizing Bell's approach threatens to
bury the measurement problem. But the measurement problem had been on
the front-burner of discussion already several decades before Bell
wrote. In fact, in his 1946 article ``On the measurement problem of
atomic physics'' (Om maalingsproblemet i atomfysikken), Bohr claims
that the existence of the quantum of action entails that a subject can
be entangled with the object of description, and that seems to destroy
the idea of the object having definite properties. He then reminds the
reader of the solution that he had articulated already in the 1920s,
i.e.\ that the notion of a communicable ``measurement result''
presupposes a classical context.\footnote{A precise notion of
  ``classical context'' emerges from Howard's (1979) explication of
  Bohr, and more recently it has been developed by Bub, Clifton,
  Dickson, Halvorson, and Landsman, among others. While Bohr does not
  use the exact phrase ``classical context'', and does not usually
  explicate such notions mathematically, he does say that classical
  concepts are needed to specify the ``conditions of description''
  (betingelser for beskrivelse), and thereby establish a boundary
  between subject and object. For a clear and modern discussion of
  these issues, see (Landsman 2006).}

\section{What is standard QM?}

I tried to draw attention to the fact that Bell accuses QM of lacking
several theoretical virtues, e.g.\ it is ``inexact'', ``ambiguous'',
``only approximately self-consistent.''  However, Butterfield requests
that we shift emphasis back to Bell's claim that QM fails to explain
macroscopic definiteness. I grant that the latter kind of problem
would make the former kind of problem rather insignificant. In
particular, if a theory says something obviously false, then what does
it matter if that theory has some vagueness? If QM is inconsistent
with the determinacy of the macroworld, then why does Bell spend so
much time insinuating that QM lacks theoretical virtues such as
exactness?

The reason is simple: unitary QM may have the virtues that Bell
extols, but he also believes that it makes false predictions. Thus,
Bell thinks that physicists make various vicious modifications to QM
in order to block its false predictions. In particular, physicists
manually intervene into QM to get the right predictions, e.g.\ by
invoking wave-function collapse, and the theory then becomes
subjective, ambiguous, inconsistent, etc.

Again, there is a certain way of manually intervening in a theory's
predictions that is definitely vicious. But I don't think that is what
is going on with QM. [Here I agree with Wallace (2014) that standard
QM does not include the eigenstate-eigenvalue link, nor the projection
postulate.] The idea that QM makes predictions about which quantities
are definite is, I think, a mistake. Instead, the ``intervention''
needed to extract predictions from QM is the innocuous intervention of
specifying what question one is asking. E.g.\ is one asking about an
object's position or about its momentum? In short, to get a
description out of QM, one needs to specify a classical context.

Recall here that the pilot-wave theory itself specifies a fixed
classical context, viz.\ the context where configuration variables are
assumed to have definite values. Bohr thought that was an appropriate
context to interpret the data in terms of ``spacetime coordination.''
However, Bohr also thought that the data could be interpreted in terms
of ``causality'', by establishing a context in which the law of
momentum conservation holds.

\section{Epistemology for entangled subjects}

Butterfield disagrees with my claim that Bell assumes that QM must be
read as a theory of two kinds of things: (classical) subjects and
(quantum) objects. In particular, in Section~\ref{us} he says: ``Bell
does not assume this.  What he \emph{does} do --- here and in other
papers --- is to contrast {[}the happy situation in classical physics
with the unhappy situation in quantum physics{]}.'' I agree that Bell
is trying to draw such a contrast, although I would describe it quite
differently. Let me use an analogy. After Darwin, one might have
contrasted the happy picture of humans as special creations of God
with the unhappy situation of humans as the result of random
evolutionary processes. Following Bohr, I think that the same kind of
contrast is applicable to the shift from classical and quantum
mechanics. In classical physics it was possible for humans to think of
themselves as capable of achieving a god's eye view of reality. But
quantum physics says: ``Sorry, you are entangled with the things you
are trying to describe. Therefore, it is impossible for you to
describe things `from the outside'.''

But now back to my claim that Bell falsely assumes that the subject
must appear in the quantum mechanical description. I grant that this
point was not clearly stated.  Let's look again, then, at what Bell
says explicitly: QM is defective because ``exactly where or when to
make [the subject-object distinction] is not prescribed.''  What
really is Bell's complaint here? I think the complaint, at root, is
against the kind of contextualism that one finds in Bohr's way of
thinking about QM, and in his epistemology in general.\footnote{A
  similar kind of contextualist view can be found in the work of Greta
  Hermann (see Crull 2017). This kind of contextualism has roots that
  go at least as far back as Hermann von Helmholtz, who stresses that
  knowledge of our physical constitution and situation should be used
  to interpret the data we receive from our senses (see Patton 2019).}
According to Bohr, QM provides a function from quantum states and
classical contexts to probability distributions. However, Bell seems
to be troubled by the idea that the correct description could depend
on the way that the subject-objection distinction is drawn, and that
the theory itself does not specify where to draw it. I would suggest
that Bell presupposes that a fundamental physical theory should
describe things in a contextless and subjectless fashion --- i.e.\ it
should give us a ``god's eye view'' or ``absolute conception'' or
``view from nowhere''. Of course, the desire for this kind of
description of reality has an esteemed pedigree in the history of
Western thought. Nonetheless, according to Bohr, the epistemological
lesson of quantum theory is that this kind of description is not
possible.

Bohr then suggests that an entangled subject can choose to ignore one
part of reality (i.e.\ some quantities) in order to engage in the
idealization that other quantities have exact values. This act, of
choosing a subset of quantities, is the specification of a classical
context; and, thus, every quantitatively exact description in quantum
physics is explicitly relativized to a context.

In this case, the situation in quantum physics is not ``happy'' like
the situation in classical physics, if by ``happy'' one means that the
context of description is essentially negligible. Classical physics
would permit that descriptions can be essentially subject-free, or at
least, that there could be lossless translation from one subject's
description to another's. However, the reason for giving up on this
kind of epistemic faith is because we now know that observers are
entangled with the things they are trying to describe. That is the
epistemological lesson of QM: humans cannot achieve the sort of
perspective-free knowledge that Spinoza and Einstein hoped they
could.\footnote{Landsman (2006) establishes a useful analogy where
  Einstein stands to Spinoza as Bohr stands to Maimonides. In Bohr's
  notes and correspondence it can be seen that he explicitly rejects
  the Spinozistic picture (see Favrholdt 2009, Chap 5).}  Of course,
Bell would have said that this kind of claim --- i.e.\ that QM has a
significant epistemological lesson --- is ``romantic'' or even
``unprofessional''. Some of Bell's admirers, e.g.\ Everettians, go for
radical ontological revision, while others, e.g.\ Bohmians, maintain
17th century meta-theoretical standards for clear and distinct
ontological description. In the latter case especially, it seems to be
assumed that the meta-theory is immune from falsification, or that
upholding old meta-theoretical presuppositions is a requirement of
intellectual integrity!  I would say, however, that physics has been
especially successful, and philosophically interesting, in cases where
it has led to revisions of meta-theory. Let us not forget that modern
epistemology arose, with Descartes, in tandem with modern physics.

\section{On reading charitably}

In Section \ref{intro}, Butterfield says that ``Bell does not shirk
his duty to try to understand what the pioneers of quantum mechanics
said about the measurement problem.'' However, the only evidence
Butterfield brings forward for this claim is that Bell mentions the
pioneers of quantum mechanics in the third paragraph of ``Subject and
Object.''  \emph{Pace} Butterfield, this paragraph provides little
evidence that Bell tried to understand what the pioneers of quantum
mechanics said and wrote. To the contrary, it shows Bell insinuating
that they were uninterested in foundational questions. He says that
``they did not wait for agreed answers before developing the theory'',
and that ``the vagueness of the postulate in no way interferes \dots
'' Being wise to Bell's rhetorical tricks, I read these sentences as
saying ``the pioneers had a philosophical problem on their hands, but
they ignored it and moved forward with deriving more empirical
predictions'' and ``they tolerated vagueness in the postulates.''
These claims are misleading at best, and they serve as the basis for
many false claims in the literature today. Far from ignoring the
foundational problems, Bohr and his colleagues discussed them, if not
exhaustively, certainly exhaustingly. So, at the very least, I
maintain that Bell is guilty of false insinuation.

More generally, Bell's writings, while an exemplar of clarity and
rigor, rarely provide careful or charitable reconstruction of the
thought of those he criticizes or uses as a foil. To take one example,
in the Appendix to ``Bertlmann's socks and the nature of reality'',
Bell takes up Bohr's reply to the EPR argument.  After one paragraph
of exposition, Bell says of Bohr's claims: ``I have very little idea
what this means.'' Interesting to know, but not exactly an exemplar of
careful scholarly engagement, nor much help in moving a constructive
dialogue forward. Imagine if you asked an undergraduate student to
analyze Quine's ``Two dogmas of empiricism'', and he reported ``I have
very little idea what this means.'' In that case, you might rightly
assume that the student either was not in a position to appreciate the
argument, or failed to put in the requisite effort.

If Bell did put a solid effort into trying to understand Bohr,
Heisenberg, or other early interpreters of quantum theory, then he
unfortunately did not document it.  Whitaker's (2016) detailed
intellectual biography does not suggest that Bell consulted Bohr's
notes and correspondence, nor that he interviewed anybody from the
inner circle in Copenhagen. What's more, Bell had no training in
metaphysics or epistemology, no experience reading and analyzing
philosophical texts, and no knowledge of the languages (German and
Danish) in which the discussions took place.\footnote{Bernstein (2011)
  points out that Bell could not read von Neumann's text until it was
  translated to English in 1955.}  The narrow, technical focus of
Bell's education might explain why he failed to read Bohr's text with
a will to understand what was being said.

Similarly, in his \emph{Omni} interview of 1988, Bell says: ``There is
this philosophy {[}i.e. Bohr's{]}, which was designed to reconcile
people to the muddle; You shouldn't strive for clarity --- that's
naive. `Muddle is sophisticated'.'' How could Bell, in so many ways an
exemplar of clarity and rigor, allow himself to make such imprecise
and groundless claims?  First of all, Bell attributes intentions to
Bohr without citing any evidence.  Even if some physicists did in fact
use Bohr's intellectual authority as an excuse for their own
intellectual laziness, why insinuate that Bohr intended that
consequence? Does Bell know something about Bohr of which neither his
contemporaries, nor his scholarly biographers, were aware? Second,
Bell insinuates that Bohr explicitly discouraged the search for
clarity, while the actual hard evidence supports the opposite
conclusion. For example, the word ``unambiguous'' (entydig) appears
hundreds of times in Bohr's writings, always in the context of
demanding that physicists give ``unambiguous, i.e. clear,
descriptions.''

To be fair, Bell said these things in an interview for a popular
science magazine, not in a scholarly monograph. So, I am not accusing
him of a moral shortcoming or of violating disciplinary
standards. However, Bell's attributions are ill-supported, and his
intellectual stature gives us no reason to repeat them, or even to
take them seriously.

\section{The discretion of the theorist}

In my original comment, I noted that even the pilot-wave theory leaves
things to the discretion of a theorist. However, this point, says
Butterfield (at the end of his Section \ref{us}): ``is neither here
nor there'' because ``what matters is to have a solution to the
measurement problem.'' I maintain that my point is relevant because
among Bell's charges against QM is that it leaves some things to the
discretion of a theorist. For example,
\begin{quote}
  \ldots the theory should be fully formulated in mathematical terms,
  with \emph{nothing} left to the discretion of the theoretical
  physicist \dots until workable approximations are needed in
  practice. (Bell 1990, p 33; emphasis mine) \end{quote} Granted, Bell
is thinking here of it being left to the discretion of a theorist when
a collapse of the wave-function occurs; and I agree that this kind of
discretion would be problematic, if it leads to inconsistent
attributions of properties. Bell's criticism is directed, then, to the
deformed ``theory'' that he had been taught as an undergraduate.

But let's think harder about whether Bell's demand makes sense. Does
it make sense to ask a theory to leave nothing to the discretion of
the theoretical physicist? Would it be possible to replace QM with a
theory like that?

In many other places, Bell suggests that the pilot-wave theory meets
his stringent demands, and so also the demand that nothing be left to
the discretion of the theoretical physicist. Butterfield clarifies by
saying that the pilot-wave theory only leaves an ``innocuous role''
for an external subject. But how are we to distinguish the innocuous
from the non-innocuous roles? I do not ask this question rhetorically,
but in all seriousness. It would behoove philosophers of physics to
raise the level of clarity about what we take to be virtues and
defects of theories. In particular, all theories leave some things to
the discretion of a theorist. Which such things are innocuous, and
which are problematic?

My own view here is that Bohr himself was arguing that classical
context belongs on the side of innocuous discretion --- even if that
destroys the illusion of a god's eye view description. Indeed, much of
Bohr's work was devoted to exploring whether contextualism is
consistent with objective description, even though it initially seems
to raise the specter of subjectivity. In any case, I mention the
pilot-wave theory to stress that every empirical theory, even the most
exact, still leaves some things to the discretion of the theorist; and
so Bell is unfair to criticize QM for having this feature.

\section{Conclusion}

\emph{De gustibus non est disputandum}. The debate we are engaged in
here is ultimately too vague to admit of any sharp resolution. It is
true that we need more careful historical and conceptual research on
Bohr, Bell, and on the history of quantum physics in general. Let's
not bury any of it, but let a hundred flowers bloom! However, even
after all the historical research is in, there will still be
disagreements of taste and style. While there are many aspects of
Bell's style that I find admirable, his philosophical heavy-handedness
is not one of them.

\bigskip \noindent {\em Acknowledgements}:--- \\
By H. Halvorson: For discussion, thanks to Jeremy Butterfield, Erik
Curiel, Ronnie Hermans, Klaas Landsman, and Chris Timpson. For
clarification of Bohr's view, I'm indebted to the work of David Favrholdt. \\
By J. Butterfield: I am very grateful to: Adam Caulton, Erik Curiel,
Sebastian De Haro, Henrique Gomes, James Ladyman, Klaas Landsman,
Ruward Mulder, Bryan Roberts, Simon Saunders, Nic Teh; and especially
to Hans Halvorson, Ronnie Hermens and Chris Timpson.

\newpage 
\section*{References}

\mbox{ }

Bassi, A.  and Ghirardi G. (2007), `The Trieste Lecture of John
Stewart Bell', {\em Journal of Physics A: Mathematical and
  Theoretical} {\bf 40}, 2919-2933, doi:10.1088/1751-8113/40/1/002

Bell, J. (1973), `Subject and Object', in Bell (1987/2004), Chapter 5.

Bell, J. (1990), `Against ``measurement''\,', \emph{Physics world}
{\bf 3}, 33--40.

Bell, J. (1987/2004), {\em Speakable and Unspeakable in Quantum
  Mechanics}, Cambridge: Cambridge University Press.

Bernstein, J. (2011), `Bell and von Neumann' arXiv:1102.2222a

Bohr, N. (1928), `Das Quantenpostulat und die neuere Entwicklung der
Atomistik', \emph{Naturwissenschaften} 16, 245--257
(1928). \url{https://doi.org/10.1007/BF01504968} Translated as as `The
quantum of action and the description of nature' in Bohr (1934),
\emph{Atomic Theory and the Description of Nature}. Cambridge
University Press.

Bohr, N. (1946), `Om maalingsproblemet i atomfysikken' in
\emph{Feststrift to N.E. Nørlund in Anledning af hans 60 Aars
  Fødselsdag den 26. Oktober 1945 fra danske Matematikere, Astronomer
  og Geodæter, Anden Del}. Ejnar Munksgaard, Copenhagen,
pp. 163--167. [Reprinted with translation `On the problem of
measurement in atomic physics' in Niels Bohr Collected Works, Vol 11,
pp 655--666. doi:10.1016/S1876-0503(08)70438-9]

Butterfield, J.  (2018), `Peaceful Coexistence: Examining Kent's
Relativistic Solution to the Quantum Measurement Problem', in {\em
  Reality and Measurement in Algebraic Quantum Theory} (Proceedings of
the 2015 Nagoya Winter Workshop), ed. M. Ozawa et al. (Springer
Proceedings Maths and Statistics, 261); pp. 277-314.
https://doi.org/10.1007/978-981-13-2487-1
http://arxiv.org/abs/1710.07844; http://philsci-archive.pitt.edu/14040

Butterfield, J.  and Marsh, B. (2019), `Non-locality and
quasiclassical reality in Kent's formulation of relativistic quantum
theory', {\em Journal of Physics: Conference Series} (DiCE 18) 1275,
012002; doi:10.1088/1742-6596/1275/1/012002 ;
http://philsci-archive.pitt.edu/15723/;
\url{https://arxiv.org/abs/1902.01294}.

Brown, H. and Wallace D. (2005), `Solving the measurement problem: de
Broglie-Bohm loses out to Everett', {\em Foundations of Physics} {\bf
  35}, 517-540.

Cator, E. and Landsman, N. (2017), `Constraints on determinism: Bell
Versus Conway–Kochen’, {\em Foundations of Physics}, {\bf 44},
781--791

Crull, E. (2017), `Greta Hermann and the relative context of
observation' in Bacciagaluppi, G. and Crull, E. (2017), \emph{Greta
  Hermann -- Between Physics and Philosophy}. Springer, pp
149--169. \url{http://doi.org/10.1007/978-94-024-0970-3_10}

Einstein, A. (1954), Letter to Bill Becker, March 20. Albert Einstein
Archives, Hebrew University Jerusalem

Favrholdt, D. (2009), \emph{Filosoffen Niels Bohr}. Informations
Forlag.

Ghirardi, G. (2014), `John Stuart Bell: recollections of a great
scientist and a great man', arxiv: 1411.1425

Halvorson, H. and Clifton R. (1999), `Maximal Beable Subalgebras of
Quantum-Mechanical Observables', {\em International Journal of
  Theoretical Physics} {\bf 38}, 2441-2484; available at:
\url{http://philsci-archive.pitt.edu/65/}

Halvorson, H. and Clifton, R. (2002), `Reconsidering Bohr’s Reply to
EPR', in {\em Non-Locality and Modality}, eds. J. Butterfield and
T. Placek: NATO Science Series (Series II: Mathematics, Physics and
Chemistry), vol 64. Springer

Hermens, R. (2019), `An operationalist perspective on setting
dependence', {\em Foundations of Physics}, {\bf 49}: 260–282

Howard, D. (1979), \emph{Complementarity and Ontology: Niels Bohr and
  the Problem of Scientific Realism in Quantum Physics}. PhD Thesis,
Boston University.

Janssen, H., (2008), `Reconstructing Reality: Environment-induced
Decoherence, the Measurement Problem, and the Emergence of
Definiteness in Quantum Mechanics', on Pittsburgh archive at
http://philsci-archive.pitt.edu/4224.

Kent, A. (2014), `Solution to the Lorentzian quantum reality problem',
{\em Physical Review A} {\bf 90}, 012107; arxiv: 1311.0249.

Kent, A. (2015), `Lorentzian quantum reality: postulates and toy
models', {\em Philosophical Transactions of the Royal Society} {\bf A
  373}, 20140241; arxiv: 1411.2957.

Kent, A. (2017), `Quantum reality via late-time photodetection', {\em
  Physical Review A} {\bf 96} 062121 ; doi:
10.1103/PhysRevA.96.062121; arxiv: 1608.04805.

Landsman, N. (2006), `When champions meet: Rethinking the
Bohr–Einstein debate’, {\em Studies in History and Philosophy of
  Modern Physics} {\bf 37}, 212--242. doi:10.1016/j.shpsb.2005.10.002

Landsman, N. (2017) {\em Foundations of Quantum Theory}, Springer:
freely downloadable from
\url{https://link.springer.com/book/10.1007/978-3-319-51777-3}

Landsman, N. (2017a), `On the notion of free will in the Free Will
Theorem’, {\em Studies in History and Philosophy of Modern Physics}
{\bf 57}, 98–103

Mann, C. and Crease, R. (1988) `John Bell particle physicist
(interview)', Omni 10(8), 84--92 and 121.

McLaughlin, B. (1992), `The rise and fall of British emergentism', in
A. Beckerman, H. Flohr and J. Kim (eds.) {\em Emergence or
  Reduction?}, Berlin: de Gruyter; reprinted in M. Bedau and
P. Humphreys (eds.) (2008), {\em Emergence: contemporary readings in
  philosophy and science}, MIT Press: Bradford Books.

Norsen, T. (2017) \emph{Foundations of Quantum Mechanics}. Springer.

Patton, L. (2019), `Perspectivalism in the development of scientific
observer-relativity' in \emph{The Emergence of Relativism},
ed. M. Kusch et al. Routledge, pp 63--78.

Putnam, H. (1992), \emph{Renewing Philosophy}. Harvard University
Press.

Schr\"odinger, E. (1926) Letter to Wilhelm Wien, August. Wien Archiv,
Deutsches Museum, Munich

Schr\"{o}dinger, E. (1935), `The Present Situation in Quantum
Mechanics': A Translation of Schr\"{o}dinger's ``cat paradox'' paper
(trans: J D. Trimmer) {\em Proceedings of the American Philosophical
  Society}, {\bf 124}, (Oct. 10, 1980), pp. 323-338; American
Philosophical Society; \url{http://www.jstor.org/stable/986572}

Seevinck, M. and Uffink, J. (2011), `Not Throwing out the Baby with
the Bathwater: Bell’s condition of local causality mathematically
`sharp and clean'.' In: Dieks D., Gonzalez W., Hartmann S., Uebel T.,
Weber M. (eds) {\em Explanation, Prediction, and Confirmation. The
  Philosophy of Science in a European Perspective}, vol 2. Springer.

von Neumann, J. (1932): {\em Mathematical Foundations of Quantum
  Mechanics}, Princeton: University Press (English translation 1955,
reprinted in the Princeton Landmarks series 1996).

Wallace, D. (2014), \emph{The Emergent Multiverse: Quantum Theory
  according to the Everett Interpretation}. Oxford University Press.

Whitaker, A. (1998), `John Bell and the most profound discovery of
science', \emph{Physics
  World}. \url{https://physicsworld.com/a/john-bell-profound-discovery-science/}

Whitaker, A. (2016), \emph{John Stewart Bell and Twentieth-Century
  Physics}. Oxford University Press.

Williams, B. (1978), \emph{Descartes: The Project of Pure
  Enquiry}. Penguin Books.

Wigner, E. (1962) `Remarks on the Mind-Body Problem', in I.J. Good
(ed.), {\em The Scientist Speculates}, London: Heinemann.

Zeh, H-D (1970), `On the interpretation of Measurement in Quantum
Theory', {\em Foundations of Physics}, {\bf 1}, 69--76.

Zeh, H-D, Joos, E. et al. (2003), {\em Decoherence and the Appearance
  of a Classical World in Quantum Theory}, second edition; Berlin:
Springer.

\section*{Appendix: What Bell says}\label{jsb}

Here, for convenience and completeness, is the beginning of Bell's
paper, as reprinted in his (1987/2004). This is the passage on which
Part I concentrates. The paper was for a symposium in honour of
Dirac. But note that the CERN preprint from September 1972 (available
at:
\url{http://cds.cern.ch/record/610096/files/CM-P00058496.pdf?version=1})
has an opening sentence deleted from the reprint, namely: `I have been
invited to contribute under this heading.' So the form of the
invitation to Bell might explain, at least in part, why he here cast
the measurement problem in the language of `subject' and `object'. It
is possible that the symposium organizers were influenced in their
choice by Bohr's frequent use of the subject/object terminology, which
traces back to more general epistemological discussions in the 19th
century.

\begin{quote} The subject-object distinction is indeed at the very
  root of the unease that many people still feel in connection with
  quantum mechanics. {\em Some} such distinction is dictated by the
  postulates of the theory, but exactly {\em where} or {\em when} to
  make it is not prescribed. Thus in the classic treatise of Dirac we
  learn the fundamental propositions: \\
  \indent \ldots any result of a measurement of a real dynamical variable is one of its eigenvalues,\\
  \indent \ldots if the measurement of the observable $\xi$ for the
  system in the state corresponding to $| x \rangle$ is made a large
  number of times, the average of all the results obtained will be
  $\langle x | \xi | x \rangle$ \dots ,\\
  \indent \ldots a measurement always causes the system to jump into
  an eigenstate of the dynamical variable that is being measured \dots
  .

  So the theory is fundamentally about the results of `measurements',
  and therefore presupposes in addition to the `system' (or object) a
  `measurer' (or subject). Now must this subject include a person? Or
  was there already some such subject-object distinction before the
  appearance of life in the universe? Were some of the natural
  processes then occurring, or occurring now in distant places, to be
  identified as `measurements' and subjected to jumps rather than to
  the Schr\"{o}dinger equation? is `measurement' something that occurs
  all at once? Are the jumps instantaneous? And so on.

  The pioneers of quantum mechanics were not unaware of these
  questions, but quite rightly did not wait for agreed answers before
  developing the theory. They were entirely justified by results. The
  vagueness of the postulates in no way interferes with the miraculous
  accuracy of the calculations. Whenever necessary a little more of
  the world can be incorporated into the object. In extremis the
  subject-object division can be put somewhere at the `macroscopic'
  level, where the practical adequacy of classical notions makes the
  precise location quantitatively unimportant. But although quantum
  mechanics can account for these classical features of the
  macroscopic world as very (very) good approximations, it cannot do
  more than that.  [footnote omitted] The snake cannot completely
  swallow itself by the tail. This awkward fact remains: the theory is
  only {\em approximately} unambiguous, only {\em approximately}
  self-consistent.

  It would be foolish to expect that the next basic development in
  theoretical physics will yield an accurate and final theory. But it
  is interesting to speculate on the possibility that a future theory
  will not be {\em intrinsically} ambiguous and approximate. Such a
  theory could not be fundamentally about `measurements', for that
  would again imply incompleteness of the system and unanalyzed
  interventions from outside. Rather it should again become possible
  to say of a system not that such and such may be {\em observed} to
  be so but that such and such {\em be} so. The theory would not be
  about `{\em observ}ables' but about `{\em be}ables'. These beables
  need not of course resemble those of, say, classical electron
  theory; but at least they should, on the macroscopic level, yield an
  image of the everyday classical world \dots
\end{quote}

\end{document}